\documentstyle[sprocl]{article}

\input{epsf}

\begin{document}

\title{PARALLELS BETWEEN QUANTUM ANTIFERROMAGNETISM AND THE STRONG
       INTERACTIONS}

\author{R. B. LAUGHLIN}

\address{Department of Physics, Stanford University, Stanford,\\
CA 94305, USA\\E-mail: rbl@large.stanford.edu}

\maketitle\abstracts{ I suggest that the great body of knowledge gained
over the past 10 years about simple spin-1/2 quantum quantum
antiferromagnets points to a connection between cuprate
superconductivity and the strong interactions.  The underlying
physical idea, which I admit to be highly speculative, is that the
phase diagram of such a magnet consists of competing ordered phases
regulated by a nearby quantum critical point.  Exactly at this critical
point the low-lying elementary excitations of the magnet are gauge
fields and particles with fractional quantum numbers analogous to the
spinon and holon excitations found in spin chains.  An arbitrarily
small distance away, however, these bind at low energy scales to make
the familiar collective modes of the ordered states into which one
renormalizes.  Vestiges of these ``parts'' of the collective modes
may be seen in conventional materials and models in high-energy
spectroscopy and inconsistencies in sum rules exactly the way
quarks are seen in particle physics.}

\begin{figure}[h]
\epsfbox{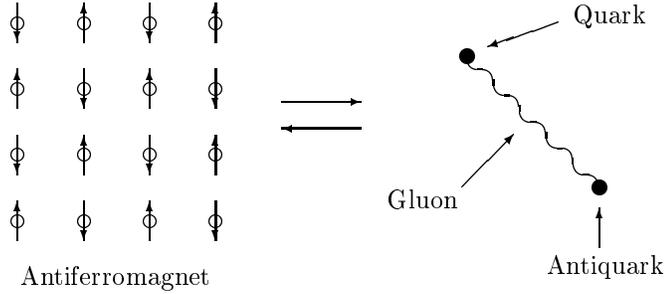}
\caption{The physical behavior of the strong interactions generated
spontaneously in simple antiferromagnets.}
\end{figure}

The premise of this article is illustrated in Fig. 1.  I wish to argue
that there are physically identifiable objects in simple Heisenberg
antiferromagnets which behave like U(1) quarks and could conceivably
be apt analogues of them.  These are not the elementary excitations
of the system in most cases but rather objects out of which the
elementary excitations are built.  It is my current belief that the
quark-like objects and the gauge fields through which they interact
are the true elementary excitations at some nearby quantum critical
point, but I shall mostly sidestep this issue and concentrate on the
physical meaningfulness of the particles in the commonly-studied cases.

The antiferromagnets in question are described by the t-J Hamiltonian

\begin{equation}
{\cal H} = P_G {\cal H}_o P_G \; \; \; ,
\end{equation}

\noindent
where

\begin{equation}
P_G = \prod_{j} \biggl\{ 1 - c^{\dag}_{j\uparrow}c^{\dag}_{j\downarrow}
c_{j\downarrow}c_{j\uparrow} \biggr\}
\end{equation}

\noindent
is the Gutzwiller projector and

\begin{equation}
{\cal H}_o =  \sum_{<j,k>} \biggl\{ -t \sum_{\sigma}
c^{\dag}_{j\sigma} c^{}_{k\sigma} +
{J\over 2}{\bf S}_j{\cdot}{\bf S}_k \biggr\} \; \; \; ,
\end{equation}

\noindent
the sum $<j,k>$ being over near-neighbor pairs of a lattice, with each
pair counted twice to maintain hermiticity.  When the dimension of the
lattice is 2 or greater, the phase diagram of this model is complex
and beyond our means to compute reliably. When the dimension of the
lattice is 1, however, there is an exact solution at
the supersymmetric point $J = 2 t$, the  ground state of which
is a nondegenerate singlet, i.e. has no order, and the elementary
excitations of which are spin-1/2, charge-0 particles known as
``spinons'' and spin-0, charge-1 particles known as ``holons''.  These
are the quark-like objects of the problem.  Their dispersion relations
are plotted in Fig. 2 \cite{bares}.  The existence of spinons and holons
is intimately connected with the lack of order in the ground state, and
is thus common in 1 dimension, where continuous symmetry breaking is
impossible, but uncommon in higher dimension, where order seems to
occur almost always. But if the higher-dimensional ground state is
{\it forced} to be disordered by means of a variational ansatz, which
is equivalent to adding a small long-range interaction to the
Hamiltonian to destabilize the order, then spinons and holons makes
sense and we obtain in 2-d the dispersion relations \cite{zou}

\begin{equation}
E_q^{spinon} = 1.6 J \sqrt{\cos^2(q_x) + \cos^2 (q_y)}
\end{equation}

\begin{equation}
E_q^{holon} = \pm 2 t \sqrt{\cos^2(q_x) + \cos^2 (q_y)}
\end{equation}

\clearpage

\begin{figure}[h]
\epsfbox{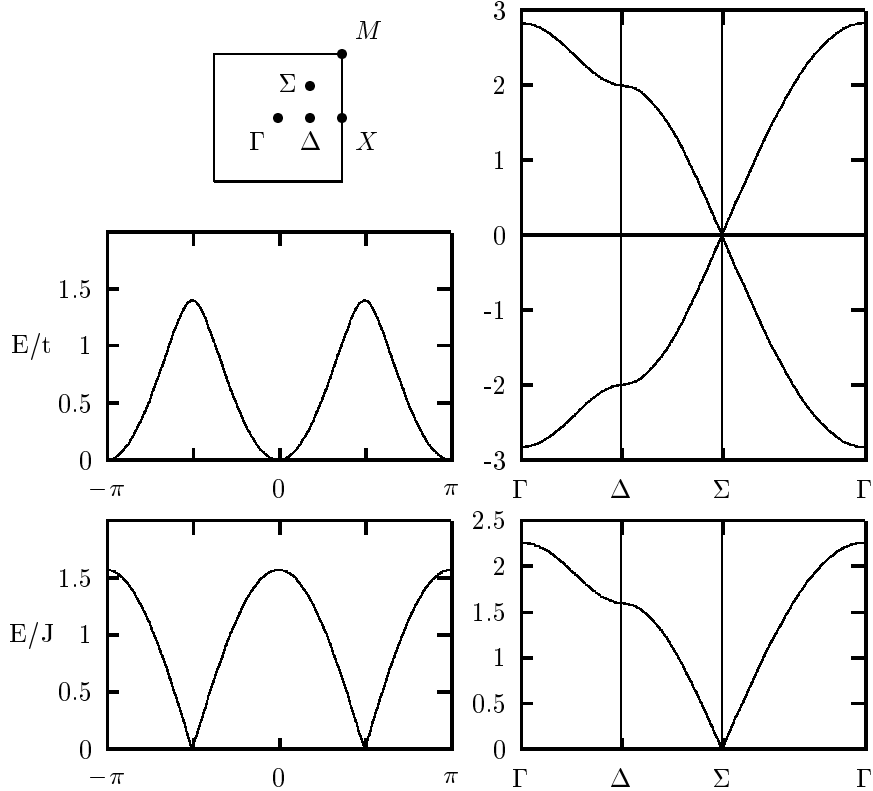}
\caption{Left: Spinon and holon dispersion relations obtained by Bares,
Blatter, and Ogata from the Bethe ansatz solution of the supersymmetric
spin chain. Right: 2-d spinon and holon dispersion relations given by
Eqs. (5) and (6). Inset: 2-d brillouin zone.}
\end{figure}

\noindent
plotted in Fig. 2. These are also implicit in the U(1) gauge theory
descriptions of the t-J model based on the commensurate flux saddle
point \cite{tikof,ioffe}. I wish now to establish that these
particles have physical meaning at intermediate energy scales even
when the system is allowed to order, giving rise to forces that bind
them at low energy scales into the well-known excitations of the
ordered phases.

In Fig. 3 I show the optical conductivity

\begin{equation}
\sigma _{xx}(\omega ) = {1\over N}{\pi \over \omega}\sum_{\alpha}
|<\alpha |j_x|0>|^2 \delta ({\hbar}\omega  - E_\alpha + E_0)
\end{equation}

\noindent
computed by Dagotto and Moreo \cite{moreo} for a single hole in
a $4 \times 4$ cluster, which is representative of such calculations.
Here $|\alpha \! >$ indicates an exact eigenstate of energy $E_\alpha$
and $j_x$ is the electric current operator

\begin{equation}
j_x = P_G \;
i{t\over \hbar}\sum_j \sum_{\sigma} \biggl\{c^{\dag}_{j\sigma}
c_{k\sigma}-c^{\dag}_{k\sigma}c_{j\sigma} \biggr\} \; P_G \; \; \; ,
\end{equation}

\noindent
where $k$ denotes the near neighbor of $j$ in the $x$-direction.
This calculation provides evidence for the existence of the holon and
measures the size of its mass.  The $f$-sum rule

\begin{equation}
\int ^\infty _0\sigma _{xx}(\omega ) d\omega  = - {\pi \over 4}
{<0|T|0>\over N}
\end{equation}

\begin{figure}
\epsfbox{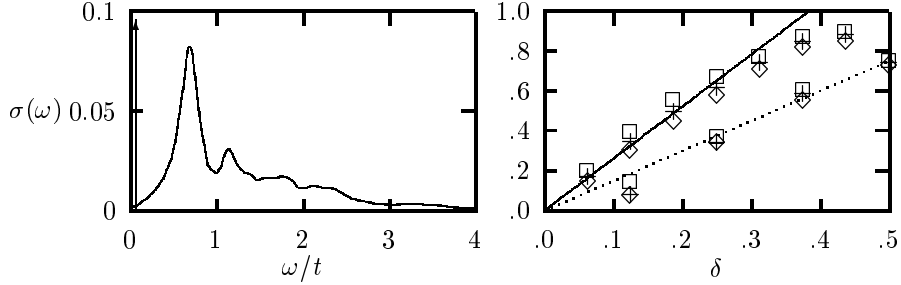}
\caption{Left: Optical conductivity computed by exact diagonalization
by Moreo and Dagotto on a $4 \times 4$ cluster for the case of
$\delta = 1/16$ and $J/t = 0.4$.  The arrow at $\omega = 0$ indicates
the Drude oscillator. Right: Total oscillator strength $- < \! T \! >
/Nt$ defined by Eq. (9) and its Drude contribution computed by exact
diagonalization on a $4 \times 4$ cluster.  The symbols $\Box$,\,$+$,
and $\Diamond$ correspond to $J/t =$ 0.1,\,0.4, and 1.0, respectively.
The solid curve is a plot of Eq. (11). The dashed curve is a guide to
the eye.}
\end{figure}

\noindent
is plotted versus doping in the lower part of the figure, as is its
``Drude'' contribution.  The width and shape of the later cannot be
computed accurately but its integrated area can.  Both sum rules are
straight lines at low doping, the slope of which {\it does not depend
on J.}  This is the behavior of a doped semiconductor.  From the
usual expression

\begin{equation}
\int ^\infty _0\sigma _{xx}(\omega ) d\omega  = {\pi \over 2}
{{\hbar}^2\over m}n \; \; \; ,
\end{equation}

\noindent
we find a mass of

\begin{equation}
m = 0.77 \; \frac{\hbar^2}{t b^2} \; \; \; ,
\end{equation}

\noindent
where b is the bond length.  This compares favorably with the
$1/\sqrt{2}$ in these same units obtained from the curvature of
Eq. (5) near its minimum and the 0.54 obtained in 1-d from the Bethe
solution.  The fact that the ``Drude'' weight is always about half
the total indicates that this particle is the carrier.  The full
sum rule

\begin{equation}
< \! T \! > = - 2.6 \; N t \delta
\end{equation}

\noindent
also agrees with Eq. (5) in equaling the $- \sqrt{8} t$ per hole
associated with the holon band minimum.  This number has no connection
to the mass in general, and is thus an additional constraint on the
band structure.

Further evidence for the existence of the holon may be found in the
electron propagator in the limit of small $J/t$. Following the notation
of Eq. (7), we define the electron propagator by

\begin{equation}
G_{q\sigma }(E) = \sum_{\alpha} \biggl\{
{|<\alpha |c^{\dag}_{q\sigma }|0>|^2\over E - E_\alpha + E_0 + i\eta }
+ {|<\alpha |c_{q\sigma }|0>|^2\over E + E_\alpha  - E_0 - i\eta }
\biggr\} \; \; \; ,
\end{equation}

\noindent
where
\begin{equation}
c_{q\sigma } = {1\over \sqrt{N}} \sum^N_j \exp (iq\cdot r_j)
c_{j\sigma } \; \; \; .
\end{equation}

\begin{figure}
\epsfbox{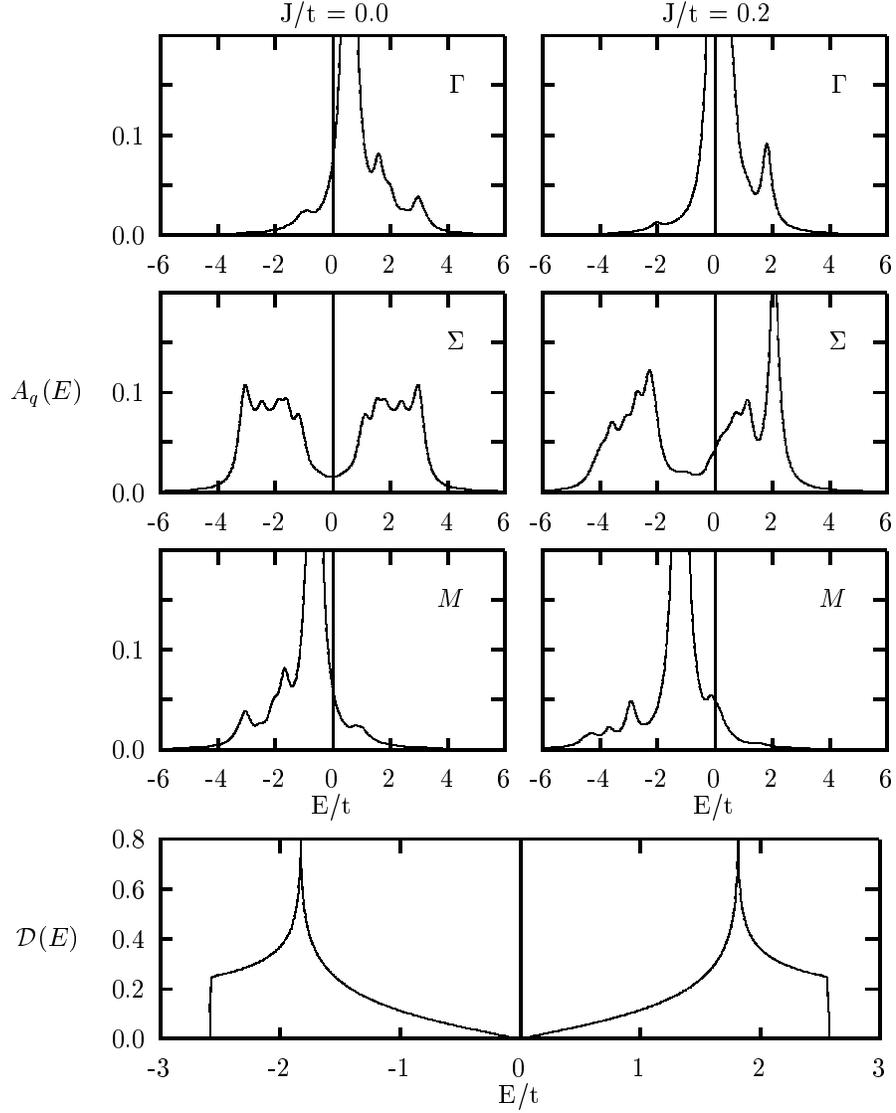}
\caption{Top: $A_q(E) = - Im G_{q \sigma}(E)$ as defined by Eq. (12),
computed by exact diagonalization on a $4 \times 4$ cluster by Dagotto.
The left and right columns correspond to $J/t = 0.0$ and $J/t = 0.2$,
respectively.  Bottom: Holon density of states defined by Eq. (14).}
\end{figure}

\noindent
In Fig. 4 I show the imaginary part of this function at half-filling
calculated by the exact diagonalization method for $J/t = 0.0$ and
0.2 by Dagotto \cite{dag}.  In either case the spectrum is a broad
continuum about 6t wide with a pronounced dip in the center and a
weight that moves from low to high energy as the momentum is
advanced from $\Gamma$ to $M$. In the $J \rightarrow 0$ limit the
broad continuum may be ascribed to the decay of the injected hole
into spinon-holon pair in the limit that the spinon is very heavy,
for then the spectrum should be the holon density of states

\begin{equation}
{\cal D}(E) = \frac{b^{2}}{2 \pi^{2}} \sum_{\lambda} \;
\int^{\pi /b}_{-\pi /b} \; \int^{\pi /b}_{-\pi /b} \;
{\delta}(E - E^{holon}_{q\lambda}) \; dq_x dq_y
\end{equation}

\noindent
weighted by a decay matrix element.  The density of states computed
from Eq. (5) is plotted in Fig. 4.  Any reasonable model will give
the motion of the weight since

\begin{equation}
< \! 0 | c_{q \sigma}^\dagger c_{q \sigma} | 0 \! > = \frac{1}{2}
\end{equation}

\begin{equation}
< \! 0 | c_{q \sigma}^\dagger \; {\cal H} \; c_{q \sigma} | 0 \! >
= < \! 0 | {\cal H} | 0 \! >
- 2 t \biggl[\frac{ < \! \vec{S}_1 \cdot \vec{S}_2 \! >}{3} -
\frac{1}{4} \biggr] \biggl\{ \cos(q_x) + \cos (q_y) \biggr\}
\end{equation}

\noindent
at half-filling, where $1$ and $2$ denote near-neighbor sites.

\begin{figure}
\epsfbox{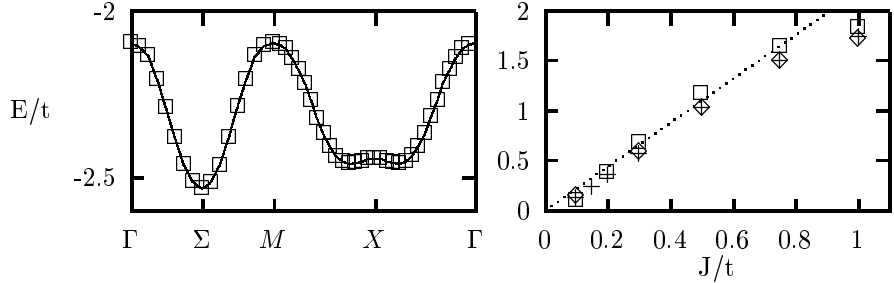}
\caption{Left: Quasiparticle dispersion relation calculated by Liu
and Manousakis using spin-wave perturbation theory for the case of
$J/t = 0.2$.  Right:  Quasiparticle bandwidth W/t calculated by Poilblanc
using exact diagonalization on clusters of various sizes. The dashed
line is a plot of Eq. (17).}
\end{figure}

The $J/t = 0.2$ curves also have a peak at low binding energy which
is the quasiparticle of the magnetic insulator.  In Fig. 5 I show
the dispersion relation of this quasiparticle found numerically
by a number of authors \cite{liu}.  It has a deep minimum at $\Sigma$
and an overall bandwidth W, the difference between the maximum and
minimum of the dispersion relation, that {\it does not depend on t.}
This width, measured in multiples of t, is plotted against $J/t$ in Fig.
5 \cite{poil}.  From the slope of the line one obtains

\begin{equation}
W = 2.2 J \; \; \; ,
\end{equation}

\noindent
or $1.6 \sqrt{2} J$, which is the spinon bandwidth given by Eq. (4).
The prefactor 1.6 in Eq. (4) has the physical significance of a
magnetic stiffness.  It causes the spinon velocity at $\Sigma$ to be
the spin-wave velocity of the ordered antiferromagnet\cite{triv}

\begin{equation}
v_s = 1.6 \frac{JB}{\hbar} \; \; \; .
\end{equation}

The quasiparticle peak is accompanied by scattering resonances.  These
cannot be seen in Fig. 4 because the sample is too small, but they may
be seen clearly in Fig. 6, which is the spectral function at $\Sigma$
for $J/t = 0.2$, calculated using spin wave perturbation
theory\cite{liu}.  The quasiparticle peak and the first two resonances
are labeled by roman numerals.  Their energies are plotted as a
function of $J/t$ in Fig. 6.  The lines through the data points
represent the formula

\begin{figure}
\epsfbox{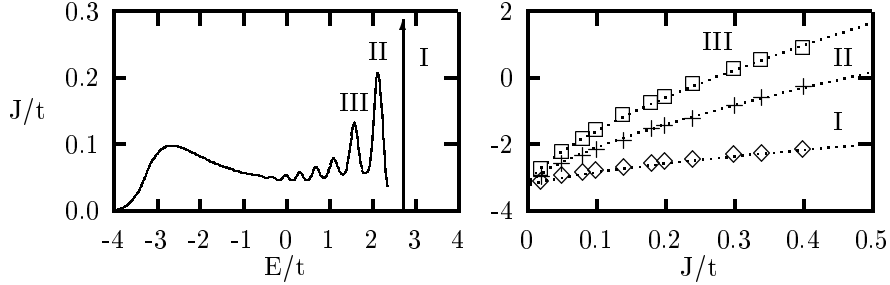}
\caption{Left: Spectral density at $\Sigma$ calculated by spin wave
perturbation theory by Liu and Manousakis for the case of $J/t = 0.1$
in the limit of large sample size.  Right: Energies of quasiparticle
(I) and first two string resonances (II and III) as a function of J/t.
The dashed lines are plots of Eq. (19).}
\end{figure}

\begin{equation}
E_n/t = - 3.28 + (J/t)^{2/3} \times  \left[\matrix{
2.03 ; n=1 \cr
5.46 ; n=2 \cr
7.81 ; n=3 \cr} \right] \; \; \; .
\end{equation}

\noindent
These energies are exactly the spectrum expected a light particle in
orbit about a heavy one, provided the attractive force between the
two is a {\it string}, i.e.  $V(r) \sim |r|$.  More precisely, the
Hamiltonian

\begin{equation}
{\cal H} = - {{\hbar}^2\over 2m} \nabla ^2 + 2.2\ J |{r\over b}| -
3.28t \; \; \; ,
\end{equation}

\noindent
where $m$ is the mass derived by the conductivity sum rule and given
explicitly by Eq. (10), has energy eigenvalues given by Eq. (19)
except for substitution (2.63, 5.54, 7.81) $\rightarrow$
(2.03, 5.46, 7.81).

These facts have the following physical interpretation.  The
quasiparticle is a bound state of a spinon and the holon analogous
to the hydrogen atom.  Its band structure tracks that of the spinon
because the spinon is ``heavier'' than the holon in the sense of
having a narrower band.  The optical sum rule, by contrast, is
sensitive to the light particle, and thus measures the holon properties.
The same thing is true in hydrogen, where the acceleration mass is
dominated by the proton but the optical properties are dominated by
the electron.  The potential binding these particles together is a
string at low doping, which means that they can never separate and
do not exist as separate entities in this limit, but already at
a doping of one hole in a $4 \times 4$ lattice, or $\delta = 1/16$,
something occurs to allow the string to break and the holon to ionize
off to become a free carrier.

The t-J Hamiltonian is formally equivalent to the Lagrangian
\cite{tikof,ioffe}

\begin{displaymath}
{\cal L} = \sum_j^N \biggl\{ \sum_\sigma f_{j \sigma}^\dagger
\biggl[ i \hbar \partial_t + \phi_j \biggr] f_{j \sigma} + b_j^\dagger
\biggl[ i \hbar \partial_t + \phi_j \biggr] b_j - \phi_j \biggr\}
\end{displaymath}

\begin{equation}
+ \sum_{<j,k>} \biggl\{ - \frac{J}{4} |\chi_{jk}|^2 + \frac{J}{2}
\chi_{jk}^* \biggl[ \sum_\sigma f_{j \sigma}^\dagger
f_{k \sigma} - \frac{2t}{J} b_j^\dagger b_k \biggr] - \frac{t^2}{J}
b_j^\dagger b_k^\dagger b_k b_j \biggr\} \; \; \; ,
\end{equation}

\noindent
where $f_{j \sigma}$ and $b_j$ are fictitious fermion and boson
operators on site j in terms of which the electron is written

\begin{equation}
c_{j \sigma} = f_{j \sigma} b_j^\dagger \; \; \; ,
\end{equation}

\noindent
$\phi_j$ is a Lagrange multiplier which when integrated out forces the
constraint

\begin{equation}
\sum_\sigma f_{j \sigma}^\dagger f_{j \sigma} + b_j^\dagger b_j = 1
\; \; \; ,
\end{equation}

\noindent
and $\chi_{jk}$ is a Hubbard-Stratonovich variable.  This is a U(1)
gauge theory to the extent that $\chi_{jk}$ may be approximated as
having a fixed length, for then the phase functions as a vector
potential along the bond $< \! j,k \! >$, the scalar potential on site
j being $\phi_j$.  This turns out to be a bad approximation for
this particular Lagrangian, but we can imagine adiabatically
transforming it into one for which $|\chi_{jk}|$ is fixed and for
which a small Maxwell term

\begin{equation}
{\cal L}_{\rm Max} =\frac{1}{g} \biggl\{ J \! \sum_{< j,k,\ell m >}
\! \chi_{jk} \chi_{k\ell} \chi_{\ell m} \chi_{mj} + \frac{1}{J}
\sum_{< j,k >} | \hbar \; \partial_t \chi_{jk} + (\phi_j - \phi_k)
\chi_{jk} |^2 \biggr\}
\end{equation}

\noindent
is added as a regulator.  Then the classical saddle point has magnetic
flux $\pi$ per plaquette, the f- and b-particles become free particles
with the relations of Eqs. (4) and (5), although with different
coefficients, and we obtain conventional lattice QED with doubled
fermions.  The limit relevant to the low-doping numerical work is
$g \rightarrow \infty$, which is strongly confining.  Thus the string
forces may be associated with confinement in strongly-coupled QED, the
antiferromagnetic order in limit may be associated with the chiral
symmetry breaking known to accompany confinement in this problem,
and the spin wave, which is both the Goldstone of the broken symmetry
and a bound pair of spinons, may be associated with the pion.

The correct appearance of a string force in the antiferromagnetically
ordered phase suggests that the unbinding of the quasiparticle seen in
Fig. 3 might indicate a first-order transition to a superconducting
phase corresponding to the coulombic phase of the gauge theory.  The
magnetic order is known to disappear at about $\delta =0.05$, which
is consistent with deconfinement by $\delta = 1/16$.  However it is
only a suggestion, for the above Lagrangian is less accurate and more
difficult to solve than the spin Hamiltonian from which it was derived,
and all the major ordering questions for the former are still
unresolved. It should be viewed not as a computational tool but as
means for understanding how the physics of the strong interactions
might materialize in a quantum antiferromagnet without being
postulated.

I wish to express special thanks to E. Dagotto for providing me his
unpublished $J \rightarrow 0$ spectral functions and to A. M. Tikofsky
for numerous helpful discussions. This work was supported primarily
by the NSF under grant No. DMR-9421888. Additional support was provided
by the Center for Materials Research at Stanford University and by NASA
Collaborative Agreement NCC 2-794.

\end{document}